\documentclass[pra,aps,twocolumn,   showpacs,superscriptaddress]{revtex4-1}

\usepackage{graphicx}
\usepackage{dcolumn}
\usepackage{bm}
\usepackage{graphicx}
\usepackage{epsfig}
\usepackage{epsf}
\usepackage{amssymb}
\usepackage{amsmath}
\usepackage{amsthm}
\usepackage{mathtools}
\usepackage{multirow}
\usepackage{enumerate}
\usepackage{enumitem}
\setenumerate{noitemsep}
\usepackage{cases}
\usepackage[colorlinks,linkcolor=dred,urlcolor=dblue,citecolor=dgreen]{hyperref}
\usepackage{pgfplots}
\usepackage{lipsum}

\definecolor{dred}{rgb}{.8,0.2,.2}
\definecolor{ddred}{rgb}{.8,0.5,.5}
\definecolor{dblue}{rgb}{.2,0.2,.8}
\definecolor{dgreen}{rgb}{.2,0.5,.2}

\newcommand{\ket}[1]{\ensuremath{|#1\rangle}}

\definecolor{mycolor1}{RGB}{25,172,142}
\definecolor{mycolor2}{RGB}{192,215,33}
\definecolor{mycolor3}{RGB}{215,75,75}
\definecolor{mycolor4}{RGB}{207,167,57}

\theoremstyle{definition}

\usepackage{pgfplots}
\pgfdeclareplotmark{orangeball}
  {\shade[draw=none,ball color=orange] (0pt,0pt) circle [radius=2pt];}

\begin{document}

\title{NMRCloudQ: A Quantum Cloud Experience on a Nuclear Magnetic Resonance Quantum Computer}

\author{Tao Xin}
\affiliation{Department of Physics, Tsinghua University, Beijing 100084, China}

\author{Shilin Huang}
\affiliation{Center for Quantum Information, Institute for Interdiscriplinary Information Sciences, Tsinghua University, Beijing 100084, China}

\author{Sirui Lu}
\affiliation{Department of Physics, Tsinghua University, Beijing 100084, China}

\author{Keren Li}
\affiliation{Department of Physics, Tsinghua University, Beijing 100084, China}

\author{Zhihuang Luo}
\affiliation{Beijing Computational Science Research Center, Beijing 100193, China}
\affiliation{Institute for Quantum Computing, University of Waterloo,
  Waterloo N2L 3G1, Ontario, Canada}

\author{Zhangqi Yin}
\affiliation{Center for Quantum Information, Institute for Interdiscriplinary Information Sciences, Tsinghua University, Beijing 100084, China}

\author{Jun Li}
\email{junli@csrc.ac.cn}
\affiliation{Beijing Computational Science Research Center, Beijing 100193, China}
\affiliation{Institute for Quantum Computing, University of Waterloo,
  Waterloo N2L 3G1, Ontario, Canada}

\author{Dawei Lu}
\email{ludw@sustc.edu.cn}
\affiliation{Department of Physics, Southern University of Science and Technology, Shenzhen 518055, China}
\affiliation{Institute for Quantum Computing, University of Waterloo,
  Waterloo N2L 3G1, Ontario, Canada}

\author{Guilu Long}
\email{gllong@tsinghua.edu.cn}
\affiliation{Department of Physics, Tsinghua University, Beijing 100084, China}

\author{Bei Zeng}
\affiliation{Department of Mathematics \& Statistics, University of Guelph, Guelph N1G 2W1, Ontario, Canada}
\affiliation{Institute for Quantum Computing, University of Waterloo, Waterloo N2L 3G1, Ontario, Canada}

\begin{abstract}
As of today, no one can tell when a universal quantum computer with thousands of logical quantum bits (qubits) will be built.  At present, most quantum computer prototypes involve less than ten individually controllable qubits, and only exist in laboratories for the sake of either the great costs of devices or professional maintenance requirements. Moreover, scientists believe that quantum computers will never replace our daily, every-minute use of classical computers, but would rather serve as a substantial addition to the classical ones when tackling some particular problems.  Due to the above two reasons, cloud-based quantum computing is anticipated to be the most useful and reachable form for public users to experience with the power of quantum. As initial attempts, IBM Q has launched influential cloud services on a superconducting quantum processor in 2016, but no other platforms has followed up yet. Here, we report our new cloud quantum computing service -- NMRCloudQ (\url{http://nmrcloudq.com/zh-hans/}), where nuclear magnetic resonance, one of the pioneer platforms with mature techniques in experimental quantum computing, plays as the role of implementing computing tasks.
Our service provides a comprehensive software environment preconfigured with a list of quantum information processing packages, and aims to be freely accessible to either amateurs that look forward to keeping pace with this quantum era or professionals that are interested in carrying out real quantum computing experiments in person.  In our current version, four qubits are already usable with in average 1.26\% single-qubit gate error rate and 1.77\% two-qubit controlled-NOT gate error rate via randomized benchmaking tests. Improved control precisions as well as a new seven-qubit processor are also in preparation and will be available later. 
\end{abstract}

\maketitle

Building universal quantum computers requires highly developed control technology on physical systems at the quantum level.
In recent years, quantum engineering technology, including preparation, manipulation, and detection of quantum systems, is undergoing rapid progress.
Quantum computing devices built on a variety of underlying physical implementations are reaching unprecedented level of control and precision in laboratories around the world.   Along with this trend,   researchers are also  making small-sized quantum simulators or processors available to the public by delivering them on the cloud \cite{IBM,Centerfor16}.  The primary benefit of cloud-based quantum computation is to enable independent parties to validate their   ideas or to   benchmark   quantum operations of interest.

In this new rising field of cloud quantum computing, the superconducting group in IBM has made first steps in 2016 \cite{IBM}. IBM provided the first commercial quantum computing service via a   web based interface called IBM Quantum Experience. The underlying quantum hardware  is a universal 5-qubit quantum computer based on superconducting transmon qubits. During its one-year service, IBM has provided a unique user experience and shed light on how to successfully maintain a server. In addition, vast amounts of data are obtained, while many of them are scientifically valuable \cite{MNWJLR17,Devitt16}.

In this work, we provide online availability of another actual quantum hardware, which is based on a nuclear magnetic resonance (NMR) spectrometer. NMR spectroscopy is arguably one of the most versatile analytic methods for investigating quantum computation and quantum control \cite{vandersypen2005nmr}. However, an NMR quantum processor (spectrometer), due to its great cost and professional maintenance requirement, is not easy for the public to gain operating experiences. In order to allow for more people, either amateurs or professionals, to embrace and more importantly participate in the tidal wave of quantum science, we launched our NMR quantum cloud computing (NMRCloudQ) service. Through NMRCloudQ, we offer direct access to a real, physical spectrometer in our lab and encourage users to explore quantum phenomena and demonstrate quantum algorithms.

In this paper, we first review some of the most important progresses in NMR  quantum computing  during the past two decades, and then briefly explain the basics of NMR quantum computing and how to access our cloud service. Finally, we   discuss the results and propose potential improvements of NMRCloudQ.

\section{NMR Quantum Computing}
Since the emergence of Shor's factoring algorithm \cite{shor1994algorithms} lit the world's fervor in pursuing quantum computers more than two decades ago, nuclear magnetic resonance (NMR) has always been one of the pioneer platforms to tackle a variety of quantum information tasks, ranging from the development of advanced control techniques \cite{vandersypen2005nmr} to implementation of quantum algorithms \cite{jones2011quantum}. Very soon after the ideas on how to build NMR quantum computers \cite{cory1997ensemble,gershenfeld1997bulk}, initial attempts towards the implementation of Grover's search algorithm \cite{jones1998implementation} and Shor's factoring algorithm \cite{vandersypen2001experimental}, and coherent control of up to 7 qubits \cite{knill2000algorithmic} have been successively realized. Such rapid progress in NMR quantum computing during the early years can be majorly attributed to the mature skills in pulse design and optimization, which are accumulated along the history of NMR spectroscopy.

On the other hand, the new application in quantum computing also brings traditional NMR community new technologies, where we would mention the gradient ascent pulse engineering (GRAPE) algorithm \cite{khaneja2005optimal} as a typical example. GRAPE techniques were firstly proposed to achieve high-fidelity controls of individual spins, as well as the entire system dynamics, to satisfy the rigid requirements of a quantum computer. At present, it has been an important pulse engineering approach in NMR spectroscopy, and is also widely used in other quantum computing architectures including election spin resonance \cite{zhang2011coherent}, nitrogen-vacancy centers in diamond \cite{waldherr2014quantum,dolde2014high}, superconducting circuits \cite{motzoi2009simple,egger2013optimized}, and ion traps \cite{nebendahl2009optimal,schindler2011experimental}. In particular, the control fidelities for single-qubit gates 99.97\% and two-qubit gates 99.5\% \cite{ryan2009randomized} demonstrated via randomized benchmarking in NMR still remain as one of the best in experimental quantum computing to date.

After two decades of the initial idea about NMR quantum computing, we launched the first cloud quantum computing service based on NMR, i.e. NMRCloudQ.

\section{Configuration of NMRCloudQ}

A rich variety of software applications for processing and analyzing NMR experimental spectra has been involved, enabling the automation
of many complex tasks in quantum information processing. With the feature-rich and sophisticated software functionality developed in the magnetic resonance community, we can reasonably conclude that most liquid-state NMR quantum computing experiments can now be automated with relative ease.

\begin{figure}[t]
\begin{center}
\includegraphics[width=\linewidth]{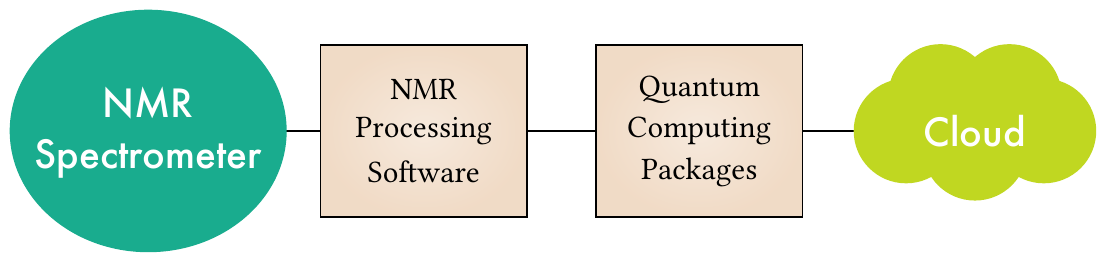}
\caption{Connection between different parts of our automated system.}
\label{cloud}
\end{center}
\end{figure}

The connections between different parts can be illustrated as shown in Fig. \ref{cloud}

\begin{figure}[t]
\begin{center}
\includegraphics[width=\linewidth]{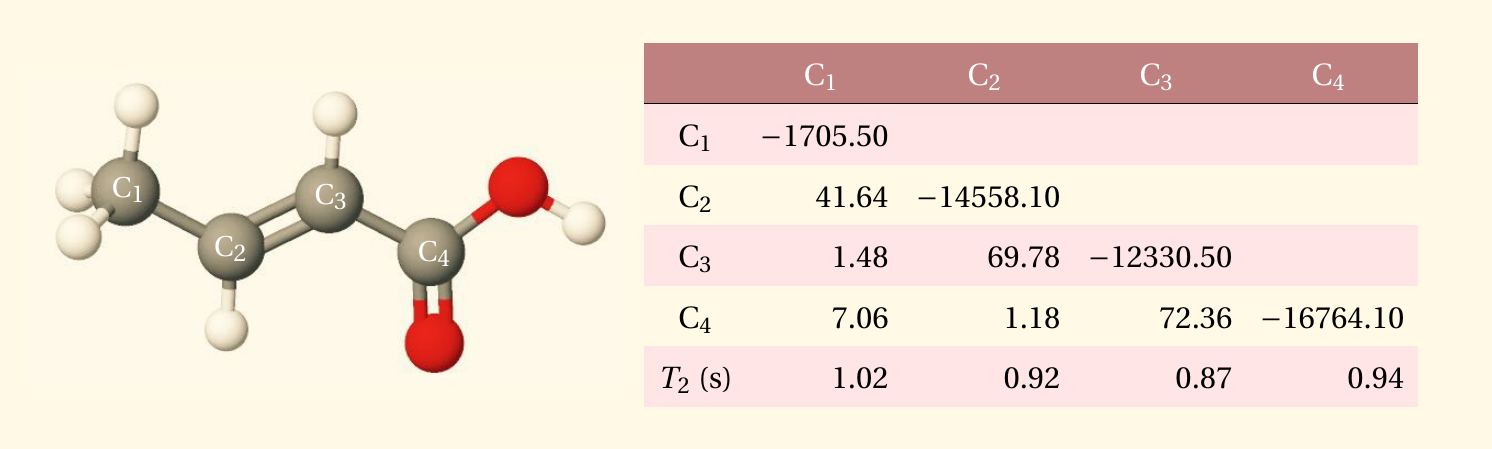}
\end{center}
\setlength{\abovecaptionskip}{-0.00cm}
\caption{Molecular structure and Hamiltonian parameters of $^{13}$C-labeled trans-crotonic acid. In experiments, C$_1$, C$_2$, C$_3$ and C$_4$ are used as a 4-qubit quantum processor. In the table, the chemical shifts and $J$-couplings (in Hz) are given as the diagonal and off-diagonal elements, respectively. The last row of the table shows the values of $T_{2}$.}\label{molecule}
\end{figure}

Our quantum processor is functionalized on a molecule $^{13}$C-labeled trans-crotonic acid (C$_4$H$_6$O$_2$) dissolved in d6-acetone. This sample contains four carbons and five protons, where the protons can be decoupled throughout experiments. Hence, we get a 4-qubit system, in which we label C$_1$ to C$_4$ as qubits 1 to 4.  See Fig. \ref{molecule} for the molecular structure and its Hamiltonian parameters.  Experiments are carried out on a Bruker AVANCE 400 MHz spectrometer in our lab at room temperature.

The Hamiltonian of the system under weak coupling approximation is written as
\begin{align}\label{Hamiltonian}
H_{S}=\sum\limits_{j=1}^4 {\pi \nu _j  } \sigma_z^j  + \sum\limits_{j < k,=1}^4 {\frac{\pi}{2}} J_{jk} \sigma_z^j \sigma_z^k,
\end{align}
where
$\nu_j$ is the Larmor frequency of the $j$th spin and $\emph{J}_{jk}$ is the $J$-coupling strength between the $j$th and the $k$th spin.

The manufacturer Bruker  provides a software $\mathtt{TopSpin}$
to control the spectrometer.
NMR experiments are routinely done with typing commands through $\mathtt{TopSpin}$. This  workstation enables users to build  their own customized experiment libraries and set up sophisticated experiments, with multiple options provided. Combining $\mathtt{TopSpin}$ with our NMR quantum computing packages, it is able to access various types of quantum information experiments remotely.

\section{NMR quantum computing Packages}

\begin{figure}[t]
\begin{center}
\includegraphics[width=\linewidth]{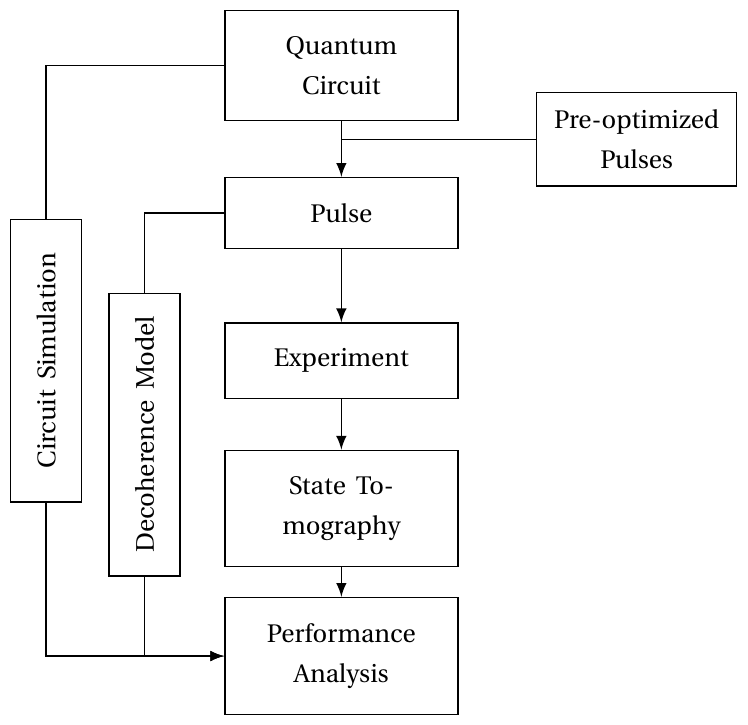}
\end{center}
\setlength{\abovecaptionskip}{-0.00cm}
\caption{Schematic workflow of our four-qubit NMR quantum computing cloud.}\label{4qubit}
\end{figure}

Generally, a quantum computing experiment contains the following steps:
\begin{enumerate}
\item[(1)] initial state preparation;
\item[(2)] implementing a quantum circuit;
\item[(3)] measurement.
\end{enumerate}
On the 4-qubit NMR quantum cloud, see Fig. \ref{4qubit}, we have stored a bunch of pre-optimized pulses. These pulses are used to implement elementary single-qubit and two-qubit quantum gates. Now let us describe the procedure of cloud computing step by step.

\subsection{Initial State Preparation}
To run a quantum algorithm, we usually need to initialize the system to a pure state, say $\ket{0}^{\otimes n}$. In NMR quantum computing, the low polarization of  the system at room temperature prevents us from generating  a genuine pure state. To see this, we write down the form of the system's equilibrium state  $\rho_{eq}$, which follows Boltzmann distribution at room temperature $T$:
\begin{equation}
\rho_{eq} = \frac{e^{-\beta H_S}}{\operatorname{Tr}\left( e^{-\beta H_S} \right)} ,
\end{equation}
where $\beta = 1/(k_B T)$ ($k_B$ is the Boltzmann constant). Since $\left\| H_S\right\|/(k_B T)  \ll 1$,
\begin{equation}
\rho_{eq}   \simeq \frac{I^{\otimes 4}}{2^4} -\beta H_S   \simeq \frac{I^{\otimes 4}}{2^4} + \epsilon \sum_{j=1}^4  \sigma_z^j,  \label{rho_eq}
\end{equation}
where Eq. (\ref{Hamiltonian}) is substituted with coupling terms omitted, and $\epsilon$ is the polarization of the $j$th spin which is in the order of $10^{-5}$.

To initialize the NMR system,  the idea of pseudo-pure state (PPS)  was introduced \cite{Chuang97,Cory98,KCL98}. PPS is a combination of the maximally mixed state and a pure state
\begin{equation}
\rho_{pps}   = (1-\eta) \frac{I^{\otimes 4}}{2^4} + \eta | 0^{\otimes 4}\rangle \langle 0^{\otimes 4} |,  \label{rho_pps}
\end{equation}
where $\eta$ is a parameter characterizing the effective purity of the PPS, and its value is in the magnitude of $\epsilon$. Although   PPS is highly mixed, the operator $I^{\otimes 4}$ does not evolve under any unitary propagator nor is it observed in NMR spectra. Hence, only the deviated part $| 0^{\otimes 4}\rangle \langle 0^{\otimes 4} |$ contributes to   experimentally observable signal.
It   appears  that PPS is  just as  suitable as the real pure state for quantum computing experiments.

We used the spatial averaging method \cite{pps99} to make the transformation $\rho_{eq} \to \rho_{pps}$. Full state tomography is then performed in order to obtain a quantitative estimation of the quality of our PPS.  We found that the fidelity between the prepared PPS and the target state is $98.77\%$. This state serves as the starting point for   subsequent computation tasks.

\subsection{Pulse Control for Executing Quantum Circuits}

Any quantum circuit can be decomposed into a sequence of elementary quantum gates. It would be ideal to provide the availability of a set of universal set of gates, such as the Clifford gates
plus $T$ gate. However, due to the structure of our molecule, non nearest-neighboring $\operatorname{CNOT}$ gates are difficult to generate since the couplings are too small. So we do not provide CNOT gates between arbitrary two qubits but only nearest-neighbor ones. All pre-computed pulses are listed in Table \ref{pulses}, while they still form a universal set of gates.

\begin{table*}
\centering
{\footnotesize
\renewcommand{\arraystretch}{1.25}\setlength{\tabcolsep}{7.5pt}
\begin{tabular}{|lrrrc|}
\hline
\multicolumn{1}{|c}{Pulse name}  & \multicolumn{1}{c}{Target gate} & \multicolumn{1}{c}{Pulse duration} & \multicolumn{1}{c}{No. of Slices} & \multicolumn{1}{c|}{Numerical fidelity} \\
 \hline
   NMRcloud\_H\_C1 & Hadamard1 & 1ms & 500 & 0.9989 \\
   NMRcloud\_H\_C2 & Hadamard2 & 1ms & 500 & 0.9992 \\
   NMRcloud\_H\_C3 & Hadamard3 & 1ms & 500 & 0.9943 \\
   NMRcloud\_H\_C4 & Hadamard4 & 1ms & 500 & 0.9943  \\
   NMRcloud\_90\_C1 & $R_x^1(\pi/2)$ & 1ms & 500 & 0.9989 \\
   NMRcloud\_90\_C2 & $R_x^2(\pi/2)$ & 1ms & 500 & 0.9986 \\
   NMRcloud\_90\_C3 & $R_x^3(\pi/2)$ & 1ms & 500 & 0.9993 \\
   NMRcloud\_90\_C4 & $R_x^4(\pi/2)$ & 1ms & 500 & 0.9991  \\
    NMRcloud\_180\_C1 & $R_x^1(\pi)$ & 1ms & 500 & 0.9992  \\
   NMRcloud\_180\_C2 & $R_x^2(\pi)$ & 1ms & 500 & 0.9991  \\
   NMRcloud\_180\_C3 & $R_x^3(\pi)$ & 1ms & 500 & 0.9994  \\
   NMRcloud\_180\_C4 & $R_x^4(\pi)$ & 1ms & 500 & 0.9991  \\
     NMRcloud\_T\_C1 & T-gate1 & 1ms & 500 & 0.9987  \\
   NMRcloud\_T\_C2 & T-gate2 & 1ms & 500 & 0.9992 \\
   NMRcloud\_T\_C3 & T-gate3 & 1ms & 500 & 0.9987 \\
   NMRcloud\_T\_C4 & T-gate4 & 1ms & 500 & 0.9990  \\
   NMRcloud\_TD\_C1 & T$^{\dagger}$-gate1 & 1ms & 500 & 0.9992  \\
   NMRcloud\_TD\_C2 & T$^{\dagger}$-gate2 & 1ms & 500 & 0.9991  \\
   NMRcloud\_TD\_C3 & T$^{\dagger}$-gate3 & 1ms & 500 & 0.9994  \\
   NMRcloud\_TD\_C4 & T$^{\dagger}$-gate4 & 1ms & 500 & 0.9991  \\
 \hline
   NMRcloud\_cnot\_C12 & $\operatorname{CNOT}_{12}$ & 15ms & 5000 & 0.9995  \\
   NMRcloud\_cnot\_C21 & $\operatorname{CNOT}_{21}$  & 15ms & 5000 & 0.9999  \\
    NMRcloud\_cnot\_C23 & $\operatorname{CNOT}_{23}$  & 7.5ms & 2500 & 0.9996  \\
   NMRcloud\_cnot\_C32 & $\operatorname{CNOT}_{32}$ & 7.5ms & 2500 & 0.9999 \\
     NMRcloud\_cnot\_C34 & $\operatorname{CNOT}_{34}$ & 9ms & 3000 & 0.9998  \\
   NMRcloud\_cnot\_C43 & $\operatorname{CNOT}_{43}$  & 9ms & 3000 & 0.9999  \\
 \hline
   NMRcloud\_swap\_C12 & $\operatorname{SWAP}_{12}$ & 40ms & 4000 & 0.9995  \\
   NMRcloud\_swap\_C23 & $\operatorname{SWAP}_{23}$   & 40ms & 4000 & 0.9998  \\
    NMRcloud\_swap\_C34 & $\operatorname{SWAP}_{34}$   & 30ms & 3000 & 0.9996  \\
 \hline
\end{tabular}
}
\label{pulses}
\caption{GRAPE pulses for universal quantum gates, including 12 single-qubit gates, 6 CNOT gates and 3 SWAP gates.}\label{pulses}
\end{table*}

Finding appropriate control pulses to implement the target quantum operation is crucial for
quantum computing experiments. A variety of numerical pulse-searching methods have been developed.  The employment of optimal control theory to solve this problem proves to be a great success. In
small-size systems,   currently the most popular method in NMR is the gradient ascent pulse engineering (GRAPE) \cite{khaneja2005optimal}.  Through the GRAPE optimal control technique,
high-accuracy control pulses can be found promptly in a 4-qubit system. In Table \ref{pulses}, the pulses were optimized to be around 99.9\% fidelity, and were also designed to be robust over a range of r.f. powers ($\pm 5$\% from the ideal power). Although not time-optimal, these pulses have smooth shapes and relatively low powers to reduce the probe heating in a long sequence.

Knowing the errors of a single gate in simulation is  not sufficient for assessing the errors present in an
actual quantum computation. In practice, the device suffers from systematic errors which may even vary between calibrations.
To get a quantitative characterization of the real  evolution  when we apply a pulse, we use randomized benchmarking (RB) protocols \cite{EAZ05}. RB has been implemented in many experimental platforms \cite{ryan2009randomized,Knill08,chow09,Paik11,rong2015experimental}. Performing RB on our device is an important task for assessing its prospects with regards to achieving high-fidelity quantum control.

We have implemented RB to characterize the average single-qubit gate and two-qubit gate error rates. Their values are 1.26\% and 1.77\%, respectively. More details, such as the average error rate for each gate listed in Table. \ref{pulses}, will come soon.

\subsection{NMR State Tomography}

NMR detection is performed on a bulk ensemble of molecules, hence the readout   is  an ensemble--averaged macroscopic measurement.  All experimental data are extracted from the free-induction decay (FID), which is the  signal induced by the precessing magnetization of   the sample  in a surrounding detection coil. FID is recorded as a time-domain signal, which consists    of a number
of oscillating waves of different frequencies, amplitudes, and phases. The signal is then subjected to Fourier
transformation, and the resulting spectral lines are fitted, yielding a set of measurement data.

For our system, the direct observables are single-coherent operators, i.e., only one qubit is in $\sigma_x$ or $\sigma_y$ while all the others are in $\sigma_z$ or I. When measuring
other operators, we apply   additional readout pulses before data acquisition.  For example, to measure a three-coherent operator $\sigma^1_x\sigma^2_y\sigma^3_y$, we need to rotate it to a single coherence $\sigma^1_x\sigma^2_z\sigma^3_z$ via a readout pulse $\exp(-i\pi/4 \sigma^2_x)\otimes \exp(-i\pi/4 \sigma^3_x)$. Therefore, to realize a full 4-qubit state reconstruction, we need a set of different readout pulses  as  listed in Table \ref{readoutpulses}, where all numerical fidelities are over 0.995.

\begin{table}
\centering
{\footnotesize
\renewcommand{\arraystretch}{1.25}\setlength{\tabcolsep}{7.5pt}
\begin{tabular}{lcrc}
\hline
\multicolumn{1}{c}{Pulse name}  & \multicolumn{1}{c}{Target gate} & \multicolumn{1}{c}{Pulse duration}  \\
 \hline
   $IIII$ & $identity$ & 0.9ms  \\
   $XXXX$ & $R_x^{1234}(\pi/2)$ & 0.9ms   \\
   $IIYY$ & $R_y^{34}(\pi/2)$ & 0.9ms  \\
   $YYXX$ & $R_y^{12}(\pi/2)R_x^{34}(\pi/2)$ & 0.9ms \\
   $IIIY$ & $R_y^4(\pi/2)$ & 0.9ms\\
   $XYXX$ & $R_x^{134}(\pi/2)R_y^{2}(\pi/2)$ & 0.9ms   \\
   $YXYI$ & $R_y^{13}(\pi/2)R_x^2(\pi/2)$ & 0.9ms   \\
   $IYXI$ & $R_y^{2}(\pi/2)R_x^3(\pi/2)$ & 0.9ms   \\
   $IIIX$ & $R_x^4(\pi/2)$ & 0.9ms  \\
   $XIYY$ & $R_x^1(\pi/2)R_y^{34}(\pi/2)$ & 0.9ms  \\
   $YXII$ & $R_y^{1}(\pi/2)R_x^2(\pi/2)$ & 0.9ms   \\
   $YYXY$ & $R_y^{124}(\pi/2)R_x^3(\pi/2)$ & 0.9ms  \\
   $XYXI$ & $R_x^{13}(\pi/2)R_y^{2}(\pi/2)$ & 0.9ms   \\
   $IIYX$ & $R_y^3(\pi/2)R_x^4(\pi/2)$ & 0.9ms  \\
   $IXIY$ & $R_x^2(\pi/2)R_y^4(\pi/2)$ & 0.9ms  \\
   $IIXI$ & $R_x^3(\pi/2)$ & 0.9ms  \\
   $IYIY$ & $R_y^{24}(\pi/2)$ & 0.9ms  \\

 \hline
\end{tabular}
}
\label{readoutpulses}
\caption{Pulse set to implement 4-qubit state tomography on NMRCloudQ.}\label{readoutpulses}
\end{table}

\subsection{Numerical Simulation}

Computations are naturally subject to  systematic  errors.  To estimate the imperfection of the optimized pulses,  we   provide three sets of data for user to evaluate the performance: (1) simulated results with the ideal quantum circuit; (2) simulated results with the optimized pulses, without and with consideration of decoherence; (3) experimentally measured results.

When user submits a quantum circuit, our server will immediately generate the corresponding pulse that implements the circuit. A simulating program then computes the   performance of the  generated pulse. Suppose the pulse is of duration $t$ and consists of $M$ slices, then the time propagator will be
\begin{equation}
U(t)= \exp(-i H_M \tau) \cdots \exp(-i H_1 \tau),
\end{equation}
where $H_m$ ($m=1,...,M$) represents the total Hamiltonian of the $m$th slice   and $\tau =t/M$ is the duration of each slice.

For our NMR quantum  computing hardware,  errors majorly stem  from decoherence. The relaxation mechanism is usually described by  an independent decoherence model, that is, the qubits undergo  uncorrelated      channels, parameterized with the set of $T_{1}^i$ and $T_{2}^i$ ($i=1,2,3,4$)  per evolution time step $\tau$.
Our cloud provides a simulator to mimic what is happening under this error model. To be concrete,   the density matrix $\hat \rho$  of the system  is, at each evolution step,
subjected to the composition of the error channels $\mathcal{E}_i$ for each qubit \cite{NC00}
\begin{equation}
  \rho \to \mathcal{E}_4 \circ \mathcal{E}_3 \circ \mathcal{E}_2 \circ \mathcal{E}_1 ( \rho),
\end{equation}
where
$\mathcal{E}_i$ represents the generalized amplitude damping and phase damping channel for the $i$th qubit. The simulated dynamics will also be shown online.
Users are advised to use such data to characterize the discrepancies between theoretical   and experimental results.

\section{Outlook and Conclusion}

The primary bottleneck that limits our 4-qubit cloud computing service is the pulse control precision. In average, the single-qubit gate error rate is 1.26\% and CNOT gate error rate is 1.77\%, which are not state-of-the-art yet \cite{ryan2009randomized}. Actually, there are a few techniques to be applied in the near future, as follows.

For the single-qubit controls, the error mainly comes from the pulse optimization procedure and inhomogeneities of the magnetic field across the sample. We used 99.5\% as the preset fidelity to optimize the pulse shape, which can be further set higher to 99.99\%. In addition, the generated pulse will also be rectified using the feedback control technique to guarantee that the pulse performance on the spectrometer is the same as the generated one \cite{lu2015experimental}. For the inhomogeneity problem, we intend to apply the r.f. selection technique \cite{ryan2009randomized}, where only a portion of the NMR sample is chosen to run tasks by destroying the other's contribution to the final signal. This can be realized by a r.f. selection sequence. The selected portion of sample feels the most homogenous magnetic field, i.e., signal-to-noise (SNR) ratio will be greatly enhanced. The drawback is that the absolute signal decreases, since only that portion but not the entire sample makes contributions to it. This drawback may lead to more repetitions of experiments and hence more running time to gain a sufficiently strong signal.

For the two-qubit controls, the error mainly comes from decoherence. The length of a CNOT gate in general requires $\mathcal{O}\left(\frac{1}{2J}\right)$ time, where $J$ is the coupling strength between two spins. In our 4-qubit processor, the largest $J$ is about 70 Hz, leading to a 7 ms CNOT gate. The $T_2^*$ time of each spin is around 400 ms, so a rough estimation of the decoherence error occurring in a CNOT gate should be more than 1\%. The optimization based on the brachistochrone approach is a potential solution, but it will not improve much. Alternatively, we intend to use another sample iodotrifluroethylene
(C$_2$F$_3$I), as the 4-qubit processor. Its couplings are around 200 Hz (roughly 2.5 ms CNOT gates), in comparison with the $T_2^*$ time 600 ms, which in principle promises much more accurate CNOT gates.

The molecule of crotonic acid can also serve as a 7-qubit quantum processor. Besides the four $^{13}$C spins, the molecule has five $^{1}$H spins, while three of them form a methyl group. The methyl group, usually treated as a combination of a spin-3/2 particle and a spin-1/2 particle, can be indeed used as one qubit after the spin-1/2 part is selected \cite{knill2000algorithmic}. Therefore, the crotonic acid can at most provide 7 qubits, and its cloud service is already in preparation.

In conclusion, cloud-based computation is an effective method for the storage and distribution of NMR data to the quantum computing community. It is our goal, in this project, to bring the state-of-the-art NMR quantum processor on-line. Our approach allows users to submit their
quantum computing tasks to an automation queue and access their data remotely, i.e., without operating the NMR spectrometer in person. At current stage, it would  be of great interest for users to run certain benchmarking circuits on the platform and demonstrate its performance.
Ongoing efforts will be devoted to developing an automated computing setup with higher control precisions and more qubits. We expect that this work can make contributions to the enhancement of understandings in NMR quantum control and quantum computing to the wide public.



\section*{Acknowledgement}

We thank Yuanzhen Chen, Luyan Sun, Xiaoting Wang, Bixue Wang, Cheng Guo, Youning Li, Jianfeng Zhang, Qi-an Fu, Yicheng Bao, Benda Xu, Rulin Tang and Yanqiao Zhu for helpful comments and discussions.
Zhangqi Yin is supported by the National Natural Science Foundation of China (Grants No. 61771278 and 11574176).
Jun Li is supported by  the National Basic Research Program of China (Grants No. 2014CB921403, No. 2016YFA0301201, No. 2014CB848700 and No. 2013CB921800), National Natural Science Foundation of China (Grants No. 11421063, No. 11534002, No. 11375167 and No. 11605005), the National Science Fund for Distinguished Young Scholars (Grant No. 11425523), and NSAF (Grant No. U1530401).
Bei Zeng acknowledges Chinese Ministry of Education under grants No. 20173080024.

\end{document}